\documentclass[aps,prc,reprint,superscriptaddress,nofootinbib,amsmath,amssymb,longbibliography]{revtex4-2}
\usepackage{adjustbox}
\usepackage[colorlinks]{hyperref} 

\begin{document}

\title{Role of exact treatment of thermal pairing in radiative strength functions of $^{161-163}$Dy nuclei}

\author{L. Tan Phuc}
\email{letanphuc2@dtu.edu.vn}
\author{N. Quang Hung}
\email{nguyenquanghung5@duytan.edu.vn}
\affiliation{Institute of Fundamental and Applied Sciences, Duy Tan University, Ho Chi Minh City 700000, Vietnam}
\affiliation{Faculty of Natural Sciences, Duy Tan University, Da Nang City 550000, Vietnam}
\author{N. Dinh Dang}
\email{dang@riken.jp}
\affiliation{Quantum Hadron Physics Laboratory, RIKEN Nishina Center for Accelerator-Based Science, 2-1 Hirosawa, Wako City, 351-0198 Saitama, Japan}
\author{L. T. Quynh Huong}
\affiliation{Department of Natural Science and Technology, University of Khanh Hoa, Nha Trang City, Khanh Hoa Province 652124, Vietnam}
\author{N. Ngoc Anh}
\affiliation{Dalat Nuclear Research Institute, Vietnam Atomic Energy Institute, 01 Nguyen Tu Luc, Dalat City 670000, Vietnam}
\author{N. Ngoc Duy}
\affiliation{Department of Physics, Sungkyunkwan University, Suwon 16419, South Korea}
\author{L. Ngoc Uyen}
\affiliation{Department of Engineering Science, University of Electro-Communications, Chofu, Tokyo 182-8585, Japan}
\author{N. Nhu Le}
\affiliation{Faculty of Physics, University of Education, Hue University, 34 Le Loi Street, Hue City 530000, Vietnam} 

\date{\today}

\begin{abstract}
The enhancement of radiative strength function (RSF) in the region of low $\gamma$-rays energy ($E_{\gamma}\leq 12$ MeV), which is caused by the pygmy dipole resonance (PDR), is treated within the phonon damping model (PDM) plus exact thermal pairing (EP) without adding any extra PDR strength function. The numerical calculations performed for $^{161-163}$Dy show that, because of the effect of EP, the EP+PDM can describe reasonably well the total RSF data in both low- and high-energy regions of $\gamma$-rays. Consequently, as compared to the conventional description within the phenomenological generalized Lorentzian (GLO) and standard Lorentzian (SLO) models, the EP+PDM calculations can eliminate at least eight free parameters. This indicates the important role of microscopic approaches towards the precise description of the RSF. In particular, temperature is found to have significant contributions to the RSF below the neutron separation energy, questioning again the validity of the Brink-Axel hypothesis in this energy region.
\end{abstract}

\maketitle
The photon or radiative strength function (RSF), defined as the average electromagnetic transition probability per unit of $\gamma$-ray energy $E_{\gamma}$ \cite{Blatt}, has an important role in the study of nuclear reaction properties such as $\gamma$-ray emission rate, reaction cross section, and/or nuclear astrophysical processes \cite{Holmes76,Kopecky,Rau97,Goriely98}. The experimental information of RSF is often divided into the low- and high-energy regions of $\gamma$-rays. In the region of high-energy $\gamma$ transitions ($S_n\leq E_{\gamma} \leq 15$ MeV, where $S_{n}$ is the neutron separation energy), the RSF is directly related to the giant electric dipole resonance (GDR) deduced from the photo-absorption cross section data, which is well understood via many ($\gamma, n$) reactions \cite{Dietrich}. However, in the region of low-energy $\gamma$ transitions ($E_{\gamma}\leq S_{n}$), the experimental RSFs are rather scarce because of the less collectivity in this region. In the late 1990s and early 2000s, the Oslo Cyclotron Group has developed an advanced technique, called the Oslo method, to simultaneously extract the nuclear level density (NLD) and RSF from the experimental primary $\gamma$-ray emitted spectra by using the light-ions induced and/or inelastic scattering reactions such as ($^{3}$He, $\alpha\gamma$), ($^{3}$He, $ ^{3}$He$'\gamma$), ($p, p'\gamma$), ($p, d\gamma$), etc \cite{Henden,Tveter,Schiller,Voinov,Agvaan,Guttorm15,Larsen,Guttormsen}. Basically, the Oslo method was developed based on the Brink-Axel hypothesis \cite{Brink,Axel}, in which the experimental primary $\gamma$-ray matrix is proportional to the product of the NLD $\rho(E^*)$ ($E^*$ denotes the excitation energy) and $\gamma$-ray transmission coefficient ${\cal T}_{X\lambda}(E_\gamma)$ or RSF $f_{X\lambda}(E_\gamma)={\cal T}_{X\lambda}(E_\gamma)/(2\pi E_\gamma^{2\lambda+1})$, where $X$ stands for the electric (E) or magnetic (M) excitation and $\lambda$ is the multipolarity. This assumption implies that the RSF does not depend on $E^*$, leaving all the dependence on $E^*$ in the NLD. Based on this assumption, many experimental NLD and RSF data below $S_n$ have been extracted and are accessible for different applications \cite{Oslo}. The Oslo method, at present, is considered as the most advanced method capable of providing reliable NLD and RSF data. In particular, a recent analysis of the RSF extracted from the photo-neutron $(\gamma, n)$ cross sections in the energy region $S_n \leq E_\gamma < 13$ MeV has shown an excellent matching between the RSF data of $^{160-164}$Dy extracted from the light-ions induced and $(\gamma, n)$ reactions at $E_\gamma = S_n$ \cite{Ren}. This finding indicated the fulfillment of the principle of detailed balance and thus, confirms the reliability of the Oslo method and data.

From the theoretical aspect, the simplest description of the RSF for the electric dipole ($E1$) transitions is given in terms of a standard Lorentzian (SLO) distribution with the energy-independent GDR width \cite{Brink}. However, this phenomenological distribution overestimates the experimental RSF data at $E_\gamma \leq S_n$ \cite{Kopecky,Carol}, indicating that the above assumption of energy-independent GDR width does not hold for the low-energy $\gamma$ transitions. In other words, the energy-dependent GDR width had better be taken into account in the theoretical calculation of RSF. In addition, recent experiments have clearly shown that the RSF in the low-energy region should have contributions from not only the GDR but also the spin-flip $M1$ or giant magnetic dipole resonance (GMDR), pygmy dipole resonance (PDR), and quadrupole $E2$ excitation \cite{Voinov,Agvaan,Larsen}. As a matter of fact, in a later model proposed by Kadmensky-Markushev-Furman (KMF) based on the Fermi-liquid theory \cite{KMF}, a functional form similar to the SLO is introduced with a GDR width, which is borrowed from the collisional damping model, depending on the excitation energy or temperature. This temperature-dependent GDR width is proposed in order to reproduce the nonzero limit of the GDR when $E_\gamma$ approaches zero. The KMF together with other improved versions such as generalized Lorentzian (GLO) \cite{Kopecky2}, enhanced GLO (EGLO) \cite{Kopecky3}, modified Lorentzian (MLO) \cite{MLO1,MLO2}, and generalized Fermi liquid (GFL) \cite{GFL} models have provided a good description of RSFs in many nuclei in the low-energy region and they become common models for the phenomenological analysis of experimental RSF data. However, these phenomenological models can not explain the mechanism of the enhancement underlying the $E1$ RSF at $E_{\gamma}\leq S_{n}$ such as the PDR. Moreover, these models have no predictive power since they are relied on a number of parameters, whose values are obtained by fitting to the existing experimental data. In this case, microscopic models are obviously preferable to phenomenological ones. 

Although a number of phenomenological RSF models have been proposed, there exists only one microscopic model, namely the quasiparticle random-phase approximation (QRPA). The latter was first proposed based on the Hartre-Fock plus BCS mean field with a Skyrme SLy4 interaction \cite{Goriely02}. The RSF for the $E1$ strength in this case is calculated by folding all the QRPA resonance strengths at the energies $E_i$, which are smoothed into a continuous line shape by the normalized Lorentzian function. However, this QRPA version still employs a temperature-independent GDR width taken from the experimental data or empirical systematics instead of being microscopically calculated from the QRPA response function. This causes some deviations of the calculated $E1$ strengths from the experimental data in the low-energy region (see e.g., Fig. 3 in Ref. \cite{Goriely02}). An improved version of the QRPA was later developed, in which the Hartree-Fock-Bogoliubov (HFB) mean field with Skyrme BSk2-7 interactions was employed and a temperature-dependent GDR width was used instead of the temperature-independent one \cite{Goriely04}. Although the $E1$ strengths obtained within this improved QRPA version are in a better agreement with the experimental data than those obtained within the previous one (see e.g., Figs. 6 and 7 in Ref. \cite{Goriely04}), this model had to employ two fitting parameters, namely $C_{ST}$ for the energy-dependent width (Eq. (18) in Ref. \cite{Goriely04}) and $\alpha$ for the temperature-dependent width (Eq. (21) in Ref. \cite{Goriely04}). In addition, two QRPA versions above have been applied to describe a large scale of $E1$ strengths in all nuclei with $8 \leq Z \leq 110$ but they have not been used to calculate the $M1$ strengths or higher. Hence, the E1 RSF obtained within these two QRPA versions cannot directly compare with the total RSF, which is the sum of all the $E1, M1, E2$, and PDR excitations. A recent extended QRPA version, which is based on an axially symmetric deformed HFB+QRPA with the finite-range Gogny D1M interaction, both $E1$ and $M1$ strengths have been calculated for a large number of nuclei \cite{Martini16,Goriely16,Goriely18} and the calculated $E1+M1$ strengths are directly compared with those extracted from the Oslo, nuclear resonance fluorescence (NRF), average resonance due (ARC), and thermal neutron capture measurements \cite{Martini16,Goriely16,Goriely18, Goriely19}. Since the D1M+QRPA version was developed with the purpose of providing a satisfactory description of available experimental data as well as being considered for practical applications in nuclear reactions and astrophysical studies, some phenomenological corrections have been employed in this model. These corrections, for example, include a broadening and a shift of the QPRA strengths and/or empirical damping of the collective motions (see e.g., Sec. 4.3 of Ref. \cite{Goriely19} and references therein).

Very recently, we have proposed a microscopic model to simultaneously describe the NLD and RSF \cite{HungPRL}. For the RSF, we employed the phonon damping model (PDM) \cite{PDM1,PDM2,PDM3}, which consistently includes the exact thermal pairing (EP) \cite{Hung09,Hung10}, in order to take into account both temperature-dependent GDR width (within the PDM) and thermal pairing (within the EP). Within the EP+PDM, the total RSF is calculated as the sum of the PDM strength functions $S_{X\lambda}(E_\gamma)$ for all the $E1, M1$, and $E2$ resonances at a given temperature $T$. In particular, it has been shown in Ref. \cite{HungPRL} that the RSFs obtained within the EP+PDM for $^{170-172}$Yb nuclei at $T=0.7$ MeV are in excellent agreement with the Oslo data at $E_\gamma \leq S_n$, indicating a violation of the Brink-Axel hypothesis. Moreover, the EP+PDM has reproduced well the enhancements of the RSF data in the region of $2.1 < E_\gamma < 3.5$ MeV in $^{171,172}$Yb, which are associated with the two-component PDR in the low-energy region. This result, which has not been reproduced in any microscopic models so far, was naturally obtained owing to the effect of EP, implying the important contribution of thermal pairing on the RSF at low energy. The goal of the present paper is to shed the light on the microscopic nature of the low-energy enhancement in the RSF data caused by the PDR. Three dysprosium isotopes $^{161,162,163}$Dy are selected for the illustration as their latest experimental RSF data have shown a strong dominance of the $E1$ strengths, whereas the spin-flip $M1$ and $E2$ resonances have negligible contributions to the total RSF \cite{Ren}. 

Following our previous work in Ref. \cite{HungPRL}, the RSF for the E1 excitation is calculated within the PDM as \footnote{The factor $\frac{\pi}{2}$ appears in Eq. \eqref{fE1} due to the transformation between the Breit-Wigner distribution used in the PDM to the Lorentzian distribution used in the global parameter fitting (see e.g., Eqs. (64) and (65) of Ref. \cite{DangEPJ16}), whereas the second term in Eq. (65) of Ref. \cite{DangEPJ16} is omitted in Eq. \ref{fE1} because its effect is negligible.}
\begin{equation}
\label{fE1}
f_{E1}(E_{\gamma},T)=\left(\frac{1}{3\pi^2 \hbar^2 c^2}\right)\frac{\pi}{2}\frac{\sigma_{E1}\Gamma_{E1}(E_{\gamma},T)S_{E1}(E_{\gamma},T)}{E_{\gamma}}~,
\end{equation}
where $\sigma_{E1}$ is the GDR cross section, $\Gamma_{E1}$ is the temperature-dependent GDR width, and $S_{E1}$ is the GDR strength function having the form as \cite{PDM2,DangPRC12}
\begin{equation}
\label{SE1}
S_{E1}(E_{\gamma},T)=\frac{1}{\pi}\frac{\gamma_{GDR}(E_{\gamma},T)}{(E_{\gamma}-E_{GDR})^2+[\gamma_{GDR}(E_{\gamma},T)]^2}~.
\end{equation}
In Eq. \eqref{SE1}, $E_{GDR}$ is the GDR energy and $\gamma_{GDR}(E_{\gamma},T)$, which is used to calculate the GDR width $\gamma_{GDR}(E_{\gamma},T)=\Gamma_{GDR}(E_{\gamma},T)/2$, is the damping of the GDR phonon given by, e.g., Eq. (4) of Ref. \cite{PDM1}. For the treatment of pairing, the EP is used and the EP+PDM formalism is well described, e.g, in Refs. \cite{DangPRC12,DangJPG13,DangEPJ16}, so we do not repeat it here. The model contains two parameters $F_1$ (for the coupling to the collective particle-hole ($ph$) configurations) and $F_2$ (for the coupling to the non-collective particle-particle ($pp$) and hole-hole ($hh$) configurations). The value of $F_1$ is fixed in order to reproduce the experimental GDR width $\Gamma_{GDR}$ at $T=0$ or that obtained from the global parameterization \cite{SJ} in case if no experimental width is available. The $F_2$ value is also selected at $T=0$ so that the GDR energy obtained from the EP+PDM does not significantly change with $T$. Another way to determine the value of $F_2$ is based on the experimental or empirical width at high $T$ in Ref. \cite{Schiller2}, that is, adjusting $F_2$ so that the calculated width at a given $T \neq 0$ is equal to the corresponding experimental or empirical value. Both $F_1$ and $F_2$ are thus temperature-independent parameters with $F_1 \ll F_2$. Within the PDM, the thermal width is described not only at the GDR energy but also at each phonon energy $\omega$, resulting in a fully microscopic temperature-dependent width, which is a realistic feature for a reliable prediction of the RSF.

    \begin{figure}
       \includegraphics[scale=0.25]{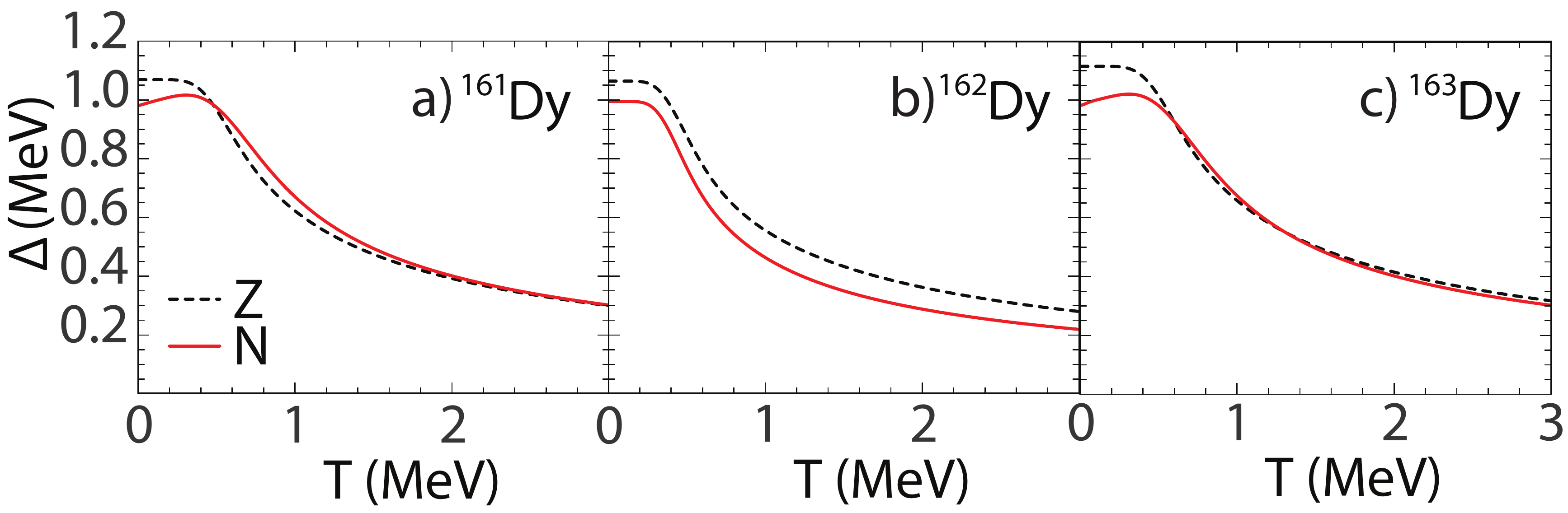}
       \caption{(Color online) Exact proton (dashed lines) and neutron (solid lines) pairing gaps as functions of $T$ for $^{161-163}$Dy.
        \label{fig1}}
    \end{figure}
    \begin{figure}
       \includegraphics[scale=0.32]{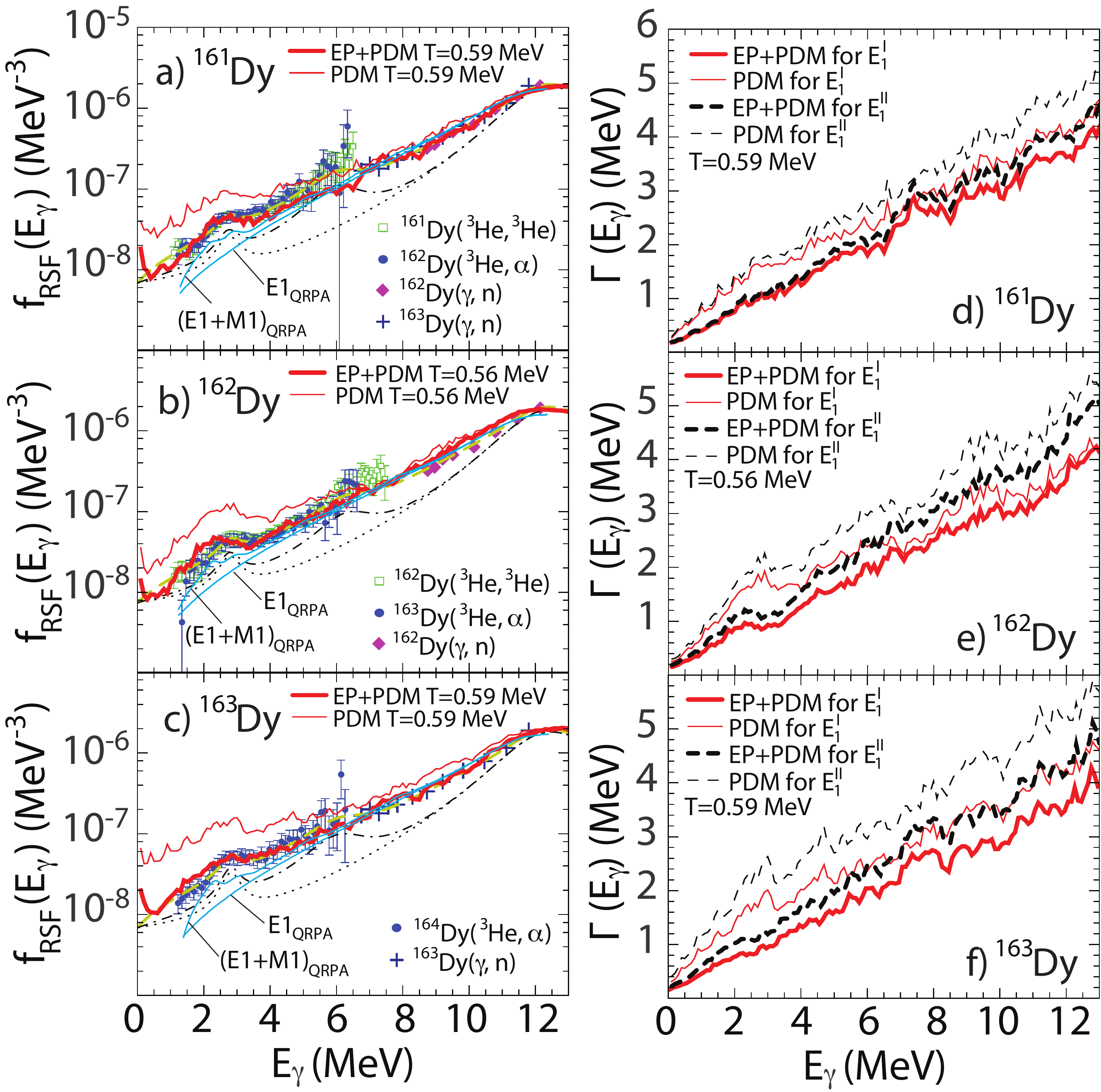}
       \caption{(Color online) [(a)-(c)] Total RSFs obtained within the PDM (thin solid lines), EP+PDM (thick solid lines) versus the QRPA RSFs for the $E1$ and $E1+M1$ excitations and the experimental data taken from Ref. \cite{Ren} for $^{161-163}$Dy. The dashed, dash-dotted, and dotted lines stand for the RSFs obtained within the phenomenological GLO-SLO models with 2 PDRs, 1 PDR, and without PDR, respectively. The QRPA $E1$ and $E1+M1$ RSFs are taken from Fig. 10 of Ref. \cite{Ren}. [(d)-(f)] GDR widths $\Gamma(E_\gamma)$ for the $E_1^I$ and $E_1^{II}$ excitations obtained within the PDM and EP+PDM for $^{161-163}$Dy.
        \label{fig2}}
    \end{figure}

\begin{table*}[t]
\caption{Resonance parameters used for the $E$1 and scissors resonance RSFs.}
\centering
\begin{tabular}{cccccccccccc}
\hline \hline
Nucleus & Method & $ E^I_1 $ & $ \sigma_{E^I_1} $ & $ \Gamma_{E^I_1} $ & $ E^{II}_1 $ & $ \sigma_{E^{II}_1} $ & $ \Gamma_{E^{II}_1} $ & $ E_{SR} $ & $ \sigma_{SR} $ & $ \Gamma_{SR} $ \\
 & & (MeV) & (mb) & (MeV) & (MeV) & (mb) & (MeV) & (MeV) & (mb) & (MeV) \\
\hline
$ ^{161} $Dy & \multicolumn{1}{l}{SJ} \cite{SJ} & 12.50 & 137.30 & 3.30 & 15.60 & 247.70 & 5.10 & 2.78 & 0.50 & 0.79 \\
 & \multicolumn{1}{l}{GLO} \cite{Ren} & 12.70 & 262.00 & 3.00 & 15.20 & 175.00 & 2.20 & 2.78 & 0.50 & 0.79 \\
 & \multicolumn{1}{l}{EP+PDM} & 12.50 & 187.70 & 3.30 & 15.60 & 231.00 & 5.10 & 2.78 & 0.50 & 0.79 \\
 \hline
$ ^{162} $Dy & \multicolumn{1}{l}{SJ} \cite{SJ} & 12.00 & 138.30 & 3.10 & 15.90 & 278.60 & 5.30 & 2.81 & 0.54 & 0.76 \\
 & \multicolumn{1}{l}{GLO} \cite{Ren} & 12.70 & 264.00 & 3.10 & 15.20 & 176.00 & 2.20 & 2.81 & 0.54 & 0.76 \\
 & \multicolumn{1}{l}{EP+PDM} & 12.00 & 198.20 & 3.10 & 15.90 & 218.70 & 5.30 & 2.81 & 0.54 & 0.76  \\
 \hline
$ ^{163} $Dy & \multicolumn{1}{l}{SJ} \cite{SJ} & 12.40 & 139.30 & 3.20 & 15.60 & 278.60 & 5.10 & 2.84 & 0.73 & 0.69 \\
 & \multicolumn{1}{l}{GLO} \cite{Ren} & 12.70 & 262.00 & 3.20 & 15.20 & 175.00 & 2.20 & 2.84 & 0.73 & 0.69 \\
 & \multicolumn{1}{l}{EP+PDM} & 12.40 & 198.20 & 3.20 & 15.60 & 239.50 & 5.10 & 2.84 & 0.73 & 0.69 \\
\hline \hline
\end{tabular} 
\label{table1}
\end{table*}

The numerical calculations are carried out for $^{161,162,163}$Dy, whose the single-particle spectra are calculated by using an axially deformed Woods-Saxon potential \cite{WS}. The parameters of the Wood-Saxon potential are chosen to be the same as those in Ref. \cite{Hung10} with the quadrupole deformation parameters $\beta_2$ equal to 0.271, 0.341, and 0.283 for $^{161}$Dy, $^{162}$Dy, and $^{163}$Dy, respectively \cite{RIPL3,Moller}. The EP calculation is performed using a truncated space consisting of 12 doubly degenerated deformed single-particle levels around the Fermi surface, where pairing has the strongest contribution. The values of the pairing interaction parameter $G$ for neutrons and protons are adjusted, as usual, to reproduce the corresponding pairing gaps at $T=0$ extracted from the odd-even mass formulas. Beyond the truncated levels, the nucleons are considered as independent particles (without pairing) and are, thus, treated by the finite-temperature independent-particle model \cite{HungPRL}. For $^{161-163}$Dy, the latest analysis of the experimental data by using the phenomenological models in Ref. \cite{Ren} has indicated that their total RSF is mostly dominated by the E1 excitations, whereas the M1 spin-flip and E2 excitations are found to be negligible. In particular, Ref. \cite{Ren} also indicated that, in addition to two GDRs described by the GLO model, the total RSFs of $^{161-163}$Dy should contain one M1 scissors resonance (SR) at $E_\gamma \approx 2.8$ MeV and two PDRs at $E_\gamma \approx$ 12.7 and 15.2 MeV, which can be both described by the SLO, namely $f_{RSF} = f_{E_1^I} + f_{E_1^{II}} + f_{PDR1} + f_{PDR2} + f_{SR}$. However, to calculate the total RSF within the EP+PDM, we need only $f_{E_1^I}$, $f_{E_1^{II}}$, and $f_{SR}$, that is, the contributions of $f_{PDR1}$ and $f_{PDR2}$ are neglected as their effect is naturally included with the PDM via the EP. These strength functions $f_{E_1^I}$, $f_{E_1^{II}}$, and $f_{SR}$ are consistently calculated within the EP+PDM by using Eq. \eqref{fE1}. The GDR and SR parameters (at $T=0$) employed in the calculations of the total RSF within the EP+PDM are given in Table \ref{table1}. Here, the GDR energies $E_1^I$ and $E_1^{II}$ and their widths are taken from the global parametrization obtained from the hydrodynamic model of Steinwedel-Jensen (SJ) in Ref. \cite{SJ}. Moreover, these GDR energies are shifted by using $E_{BW}^2=E_{L}^2-(\Gamma/2)^2$ according to the transform between the Lorentzian (L) and Breit-Wigner (BW) distributions \cite{DangEPJ16}. In the present work, to obtain systematic calculations for all nuclei, even when the experimental values for $E_{GDR}$ are not available, we employ the global values $E_L$ taken from Ref. \cite{SJ}. The parameters of the SR are kept unchanged as in Table IV of Ref. \cite{Ren}. As for the GDR cross sections $\sigma_{E_1^I}$ and $\sigma_{E_1^{II}}$ used in our calculations, they are directly calculated from the EP+PDM strength functions $S_{E_1}(E_\gamma)$ at $T=0$, namely $\sigma(E_{\gamma})=C \times S_{E1}(E_{\gamma})E_{\gamma}$ \cite{DangEPJ16}. Here, $C$ is a normalization factor to ensure the fulfillment of the GDR sum rule $C=TRK(1+\kappa)/\int S_{E1}(E_{\gamma})E_{\gamma}dE_{\gamma}$, where $TRK=60NZ/A$ (MeV mb) is the Thomas-Reich-Kuhn sum rule. The enhancement factor $\kappa = 0.5-0.7$ is caused by the meson-exchange forces. In the present work, we choose the average value $\kappa=0.6$. The EP+PDM strength function $ S_{E1}(E_{\gamma})=S_{E_1^I}(E_{\gamma})+S_{E_1^{II}}(E_{\gamma})$ is used for the axially deformed nuclei. The satisfaction of GDR sum rule leads to the total integrated cross section $\sigma=\int\sigma(E_{\gamma})dE_{\gamma}=TRK(1+\kappa)$. The cross sections of two GDR components, which are referred to as the EP+PDM cross sections, are calculated based on an assumption that the total area is equal to the integrated cross section of the two GDR components in an axially deformed nucleus and thus $\sigma_{E_1}\Gamma_{E_1}=\sigma_{E_1^I}\Gamma_{E_1^I}+\sigma_{E_1^{II}}\Gamma_{E_1^{II}}$ \cite{Berman}, where $\dfrac{\sigma_{E_1^{II}}\Gamma_{E_1^{II}}}{\sigma_{E_1^I}\Gamma_{E_1^I}}=1.9\pm0.07$ \cite{Gurevich}. The obtained EP+PDM cross sections shown in Table \ref{table1} are within those taken from the GLO fitting \cite{Ren} and global parametrization of SJ \cite{SJ}.

Figure \ref{fig1} shows the neutron and proton pairing gaps as functions of $T$ obtained within the EP for $^{161-163} $Dy. These pairing gaps are calculated within the canonical ensemble in order to mimic the mean-field pairing ones \cite{DangPRC12}. It is clear to see in this figure that the exact neutron and proton gaps decrease with increasing $T$ and remain finite even at $T=$ 3 MeV. This feature of the EP gap is well-known for finite nuclei \cite{Martin,Moretto72,HungROPP}. In addition, the exact neutron gaps of odd $^{161,163} $Dy nuclei [Figs. \ref{fig1}(a) and \ref{fig1}(c)] slightly increase at low $T \approx 0.2$ MeV due to the weakening of the blocking effect at finite $T$ \cite{HungPRC16}. Figure \ref{fig2} depicts the total RSFs obtained within the PDM with and without EP in Fig. \ref{fig1} versus the experimental data in Ref. \cite{Ren} as well as those obtained within the microscopic D1M+QRPA ($E1$ and $E1+M1$) and phenomenological GLO-SLO models. Here, the QRPA RSFs for $E1$ and $E1+M1$ are taken from Fig. 10 of Ref. \cite{Ren}. The GLO-SLO is divided into three cases, namely with 2 PDRs, with 1 PDR, and without PDR. It is seen in Fig. \ref{fig2} that the total RSFs obtained within the GLO-SLO without and with 1 PDR significantly underestimate the experimental data. The PDM calculations without EP and at finite $T$ overestimate the experimental RSF, whereas due to the effect of EP, the RSFs obtained within the EP+PDM agree reasonably well with the measured data in both $E_\gamma$ regions below and above $S_n$. The QRPA $E1$ RSF agrees with the experimental data at $E_\gamma \geq S_n$ only, while adding the $M1$ excitation to the $E1$ REF significantly improves the overall description of the QRPA. This result of EP+PDM RSF can be explained as follows. The PDM itself (without EP) includes all the thermal excitations of not only $ph$ but also $pp$ and $hh$ configurations, which appear at finite temperature. These correlations enhance the RSF at low energy ($0 < E_\gamma < 6$) MeV as seen by the thin solid lines in Figs. \ref{fig2}[(a)-(c)]. When the EP is included in the PDM, the E1 strength is enhanced, but at $T = 0$ only (see e.g., Ref. \cite{DangJPG13}). With increasing $T$, the EP+PDM E1 strength is slightly reduced as compared to the PDM one but this difference between the EP+PDM and PDM E1 strengths is relatively small, which insignificantly changes the RSF. However, the EP+PDM width $\Gamma_{E_1}(E_\gamma,T)$ is significantly reduced as compared to that given by the PDM without EP (see Figs. \ref{fig2}[(d)-(f)]). This result is similar to that reported in Refs. \cite{PDM3} and \cite{DangPRC12}. This decrease in the width leads to smaller values of the EP+PDM RSFs, which eventually match the experimental RSF, as seen in Figs. \ref{fig2}[(a)-(c)]. Hence, pairing has an important role in this case, namely reducing the width of the E1 strength distributions at low $T$ or $E_\gamma$. In other words, pairing makes the collective dipole excitations more stable than the non-paring case.

The temperatures at which the EP+PDM RSFs agree with the experimental data are found to be 0.56 MeV for $^{162}$Dy and 0.59 MeV for both $^{161}$Dy, and $^{163}$Dy. These temperatures are close to the $T_{CT}$ values obtained independently from the analysis of the corresponding NLDs using the constant-temperature (CT) model as well as those which provide the right pairing strength and lead to the good description of nuclear structure. The most interesting result here is that while the GLO-SLO analysis for the RSF needs to add two PDRs on top of the GDRs, our EP+PDM as well the QRPA do not need to include any extra strength function. In the other words, the presence of the PDRs in the RSFs of dysprosium isotopes can be microscopically explained by the effect of ground-state pairing within the QRPA as well as the effect of EP within the EP+PDM. Moreover, within the PDM and/or EP+PDM, the RSF at $E_\gamma \leq S_n$ increases with $T$, whereas its value at $E_\gamma > S_n$ is insignificantly changed as $T$ varies, similar to that reported in Ref. \cite{HungPRL}. The above results indicate that the Brink-Axel hypothesis is unlikely to hold, which is in line with our previous study for $^{170-172}$Yb in Ref. \cite{HungPRL}.

    \begin{figure}
       \includegraphics[scale=0.5]{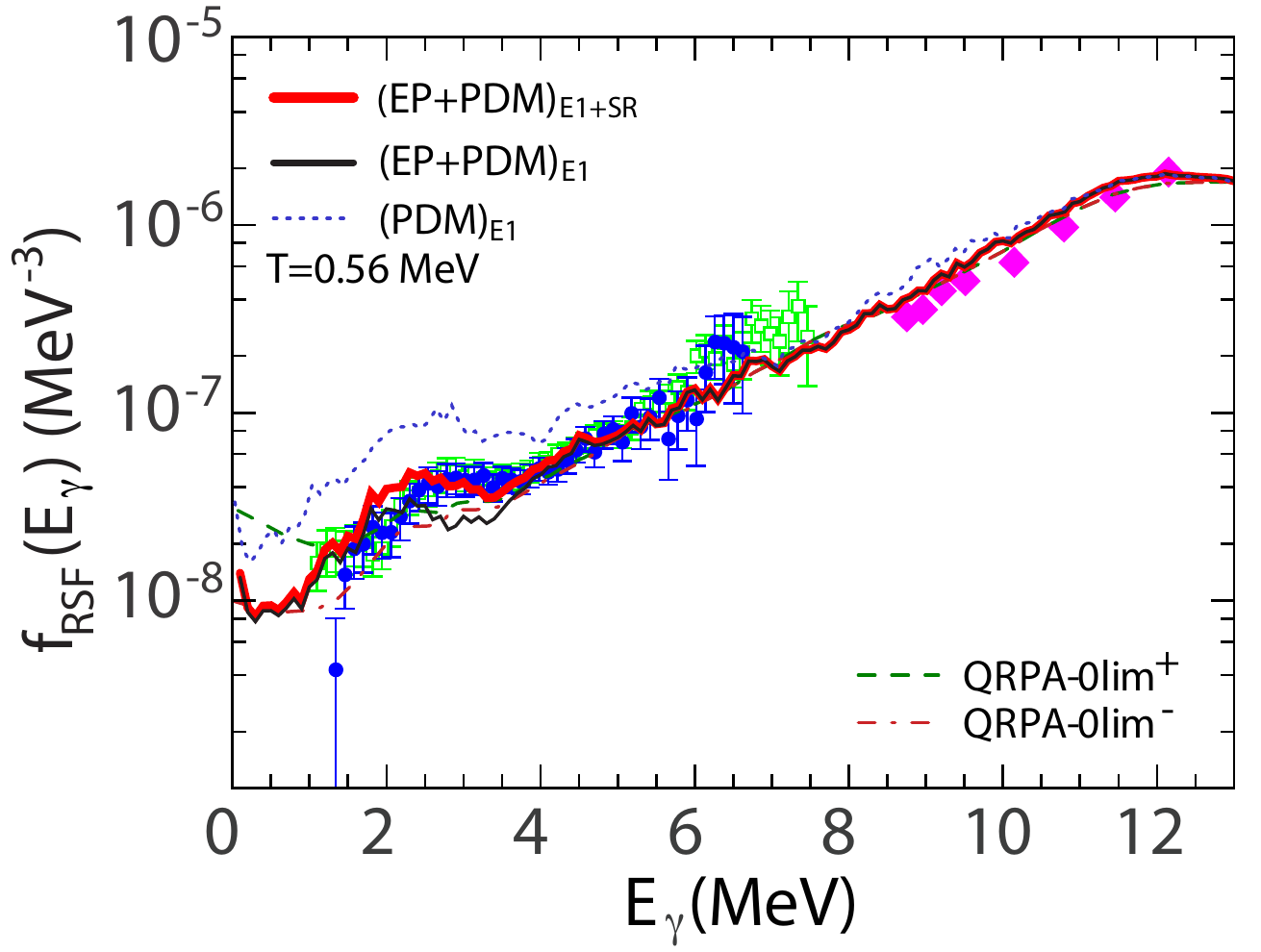}
       \caption{(Color online) RSFs obtained within the EP+PDM for the E1 (thin solid line) and E1+SR (thick solid line) excitations versus the QRPA RSFs for the lower (QRPA+0lim$^-$) (dash dotted line) and upper (QRPA+0lim$^+$) (dashed line) limits taken from Refs. \cite{Goriely18,Goriely19} for $^{162}$Dy. The RSF obtained within the PDM (dotted line) for the E1 excitation only is also shown for comparison. Experimental RSF data are the same as those in Fig. \ref{fig2}(b).
        \label{fig3}}
    \end{figure}

Since the EP+PDM is able to describe the RSF without the need of adding any extra PDR as discussed above, we have reduced at least eight free parameters as compared to the phenomenological description within the GLO-SLO model, significantly reducing the uncertainty in the theoretical prediction. This is the advantage of using a microscopic model rather than the phenomenological ones. In Fig. \ref{fig3}, we make a comparison for the RSFs of $^{162}$Dy obtained within our EP+PDM (with and without using the SR) and the results of the D1M+QRPA+0lim$^+$ and D1M+QRPA+0lim$^-$ calculations taken from Refs. \cite{Goriely18,Goriely19} as well as the PDM calculation for the E1 excitation only. This figure clearly shows that although the PDM E1 RSF overestimates the experimental data, it has reproduced a part of the SR around $E_\gamma = 1-4$ MeV. This is due to the thermal couplings of the $ph, pp$, and $hh$ excitations within the PDM. As the EP is taken into account, the EP+PDM E1 RSF decreases to match nicely with the experimental SR around $E_\gamma = 1-2$ MeV, while it underestimates the experimental SR at $E_\gamma = 2-4$ MeV. Hence, adding the SR excitation, the full EP+PDM RSF for the E1+SR agrees well with the experimental RSF in the whole energy range. Meanwhile, although both QRPA+0lim$^-$ and QRPA+0lim$^+$ RSF are in overall agreement with the measured RSF data, they underestimate the experimental SR. The reason might be due to the thermal effect caused by the $ph$, $pp$, and $hh$ couplings, which are not taken into account within the above QRPA calculations. A similar result is also seen in $^{161,163}$Dy but the contribution of the EP+PDM peak to the SR in $^{163}$Dy is weaker than that in $^{161,162}$Dy.

In conclusion, the present paper has applied the EP+PDM developed in Ref. \cite{HungPRL} to microscopically study the total RSFs in $^{161-163}$Dy nuclei. The results obtained show that, due to the effect of EP, the EP+PDM can describe reasonably well the RSF data in both low (below $S_n$) and high-energy (above $S_n$) regions without the need of using any extra strength function. As a result, at least eight free parameters have been reduced within the EP+PDM calculations as compared to the description by the phenomenological GLO-SLO model. Temperature is found to have a significant effect on the RSF at the low energy $E_\gamma \leq S_n$, whereas it does not change much the RSF in the high-energy one $E_\gamma > S_n$, questioning again the validity of the Brink-Axel hypothesis. In addition, due to the effects of EP and couplings of all $ph$, $pp$, and $hh$ configurations, the EP+PDM can also partially reproduce the scissors resonance in $^{161-163}$Dy nucleus at low $E_\gamma$ without the need of including a SR strength function in the RSF. These findings indicate the importance of EP and couplings to non-collective $pp$ and $hh$ configurations at finite temperature in the microscopic description of total RSF in excited nuclei. 

L.T.P. and N.Q.H. acknowledge the support of RIKEN, where this work was initialed. This work is funded by the Ministry of Science and Technology (MOST) of Vietnam under the Program of Development in Physics towards 2020 (Grant No. DTDLCN.02/19) and the National Foundation for Science and Technology Development (NAFOSTED) of Vietnam (Grant No. 103.04-2019.371). N.N.D thanks the support of the National Research Foundation of Korea (NRF) through Grant No. NRT-2020R10101006029.

\end{document}